\documentclass[prl,twocolumn,amsmath,amssymb,showpacs,superscriptaddress]{revtex4}
\usepackage{graphicx,epstopdf,bm}

\newcommand{\Fkt}[1]{\,\mathsf {#1}}
\ifx\Tr\renewcommand{\Tr}{\Fkt{Tr}}
\else\newcommand{\Tr}{\Fkt{Tr}}
\fi

\begin{document}

\title{Coherent Control of Bond Making}

\author{Liat Levin}
\thanks{These authors contributed equally.}
\affiliation{The Shirlee Jacobs Femtosecond Laser Research Laboratory,
  Schulich Faculty of Chemistry, Technion-Israel Institute of
  Technology, Haifa 32000, Israel}

\author{Wojciech Skomorowski}
\thanks{These authors contributed equally.}
\affiliation{Theoretische Physik, Universit\"at Kassel,
  Heinrich-Plett-Stra{\ss}e 40, 34132 Kassel, Germany
}

\author{Leonid Rybak}
\affiliation{The Shirlee Jacobs Femtosecond Laser Research Laboratory,
  Schulich Faculty of Chemistry, Technion-Israel Institute of
  Technology, Haifa 32000, Israel}

\author{Ronnie Kosloff}
\affiliation{Fritz Haber Research Centre and The
    Department of Physical Chemistry, Hebrew University, Jerusalem
    91904, Israel}

\author{Christiane P. Koch}
\affiliation{Theoretische Physik, Universit\"at Kassel,
  Heinrich-Plett-Stra{\ss}e 40, 34132 Kassel, Germany
}

\author{Zohar Amitay}
\affiliation{The Shirlee Jacobs Femtosecond Laser Research Laboratory,
  Schulich Faculty of Chemistry, Technion-Israel Institute of
  Technology, Haifa 32000, Israel}

\begin{abstract}
We demonstrate
for the first time
coherent control of bond making, a milestone on the way to coherent
control of photo-induced bimolecular chemical reactions.
In  strong-field multiphoton
femtosecond photoassociation experiments, we find
the yield of detected magnesium dimer molecules to be
enhanced for positively chirped pulses and suppressed for negatively
chirped pulses. Our \textit{ab initio} model shows that
control is achieved by purification via Franck-Condon filtering
combined with chirp-dependent Raman transitions.
Experimental closed-loop phase optimization using a learning algorithm
yields an improved pulse that utilizes
vibrational coherent dynamics in addition to  chirp-dependent Raman
transitions.
Our results show that coherent control of binary photo-reactions is feasible
even under thermal conditions.
\end{abstract}
\date{\today}
\maketitle

A long-standing yet unrealized dream since the early days of coherent
control, about 30 years ago, is the coherent control of
photo-induced bimolecular chemical reactions~\cite{tannor1,RonnieDancing89}.
Realizing this dream will create a new type of photochemistry
with selective control of yields and branching
ratios~\cite{RiceBook,ShapiroBook}.
Shaped femtosecond laser pulses act there
as special "photo-catalysts"
with a first pulse inducing and controlling the formation of a
chemical bond, and a second time-delayed pulse breaking the
desired bonds within the generated molecule.
The second step,
photodissociation into target channels with the desired
branching ratios, has been demonstrated early
on~\cite{GordonARPC97,BrixnerCPC03,DantusCR04,WollenhauptARPC05,SFB450book,LevisScience2001,RabitzScience2000},
once femtosecond lasers and
pulse shaping technology became available.
On the other hand, and in striking contrast, no study has
previously demonstrated coherent control of bond making.
The photo-induced creation of a chemical bond between the
colliding reactants, also termed photoassociation, using femtosecond
laser pulses has proven to be much more
challenging~\cite{MarvetCPL95,SalzmannPRL08,NuernbergerPNAS10,RybakFaraday11,RybakPRL11}.
Particularly at high temperature, a typical situation for chemical
reactions, the
starting point for photoassociation is rather unfavorable to coherent
control since many scattering states are incoherently populated. A
necessary requirement for the coherent control of photoassociation is
thus preparation of quantum states with some coherence. Key are
vibrational coherences in the desired bond. 
As we have previously demonstrated with two-photon
femtosecond photoassociation of hot magnesium
atoms~\cite{RybakPRL11,AmaranJCP13},
such coherences can be generated by Franck-Condon filtering of
quantum correlated states, exploiting correlations
between rotational and translational motion in
the initial incoherent thermal ensemble.
These coherences should be amenable to coherent  control.

Here we demonstrate coherent control of bond-making in strong-field
multiphoton femtosecond photoassociation (PA) of hot magnesium atoms.
Our experimental results show the PA yield of detected Mg$_{2}$
molecules to be coherently controlled by
linearly chirped pulses: The yield is strongly enhanced, compared to
an unshaped transform-limited pulse, by positively chirping the
pulses, and significantly suppressed for negatively chirped pulses.
The measured PA yield is further
enhanced by performing a closed-loop phase optimization  of
the best positively chirped pulse, using a genetic algorithm.
Our \textit{ab initio} model reveals the control mechanism
to include purification via Franck-Condon filtering of collision
energies and partial waves, chirp-dependent coherent Raman transitions,
and vibrational coherent molecular dynamics.
Our results prove that coherent control of binary photo-reactions is
feasible even under thermal conditions.

\begin{figure}[tb]
  \includegraphics[width=\linewidth]{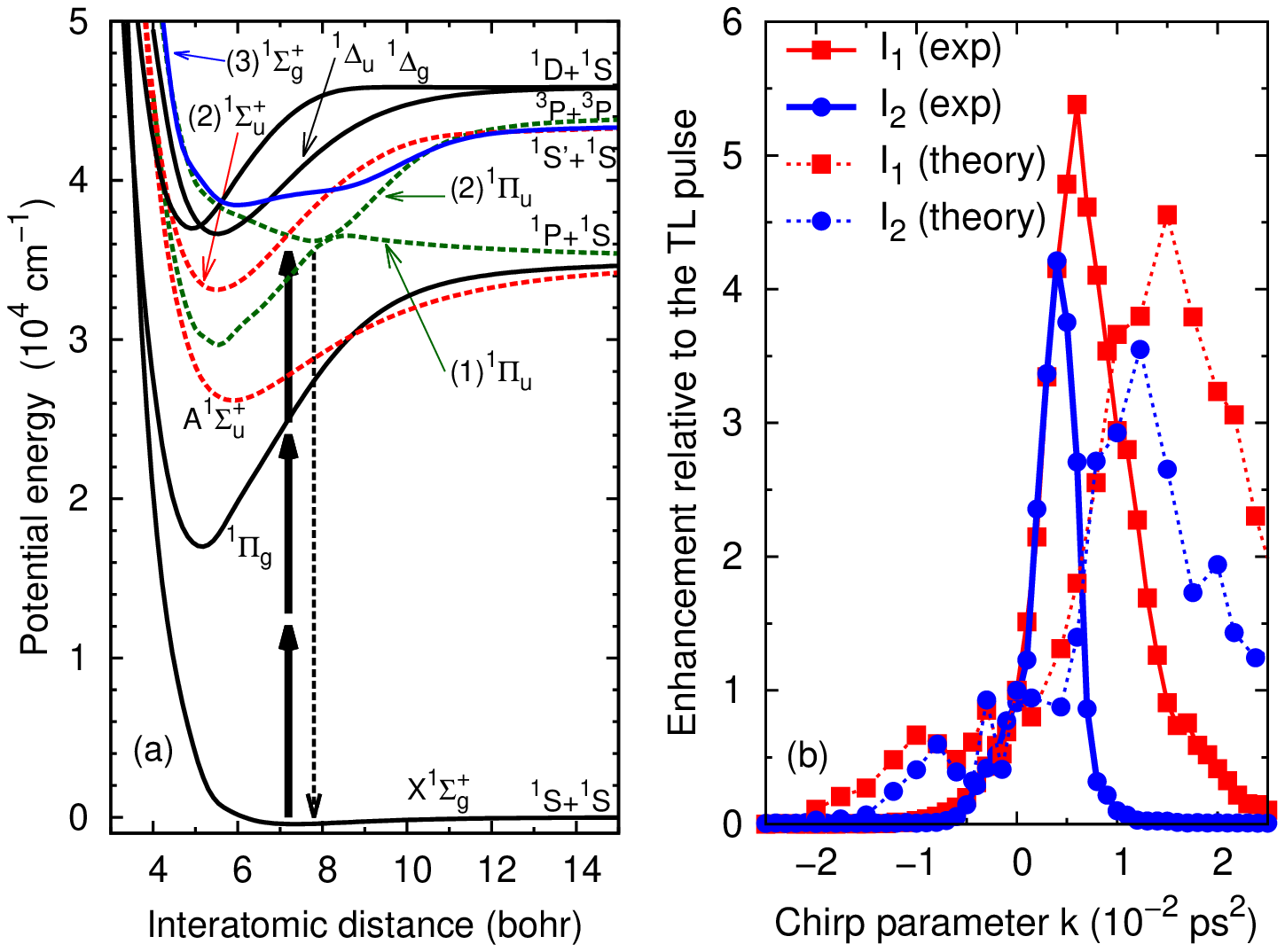}
  \caption{(a) The bond-making photoassociation process involves
    excitation of pairs of atoms
    into bound levels of electronically excited states (bold arrows)
    which are monitored by UV emission (dotted arrow).
    (b) Emitted UV intensity $I_{UV}$ (see text)
    as a function of the linear chirp parameter
    (squares: $I_{TL}=7.2\times 10^{12}\,$W/cm$^2$;
    circles: $I_{TL}=5\times10^{12}\,$W/cm$^2$).
  }
  \label{fig:scheme}
\end{figure}
The bond-making excitation scheme for the free-to-bound PA
process, Mg$+$Mg$+h\nu$$\rightarrow$Mg$_{2}^{*}$,
is shown in Fig.~\ref{fig:scheme}(a):
Pairs of magnesium atoms, part of an ensemble held at a temperature of
1000$\,$K, collide in the $X^{1}\Sigma_{g}^{+}$ ground electronic  state
and are photoassociated via multiphoton transitions by
an intense phase-shaped near-infrared femtosecond pulse.
The laser pulse is of linear polarization,
840$\,$nm central wavelength, 13$\,$nm bandwidth, and
70$\,$fs transform-limited (TL) duration.
For the linearly chirped pulses, the spectral phase of the pulse is
of the form $\Phi(\omega) = \frac{1}{2} k (\omega-\omega_{0})^{2}$
with $\omega_{0}$ the central frequency and $k$ the linear chirp parameter.
The PA process involves a broadband free-to-bound
non-resonant two-photon transition
from the $X^{1}\Sigma_{g}^{+}$ state to the excited $(1)^{1}\Pi_{g}$ state
as well as strong-field dynamics and resonant dipole transitions
between the latter and higher-lying electronically excited states.
The pulse shape controls the PA dynamics and
the final populations of the various electronically excited states.
Within this manifold,  bound rovibrational levels of
the $A^{1}\Sigma_{u}^{+}$, $(2)^1\Sigma_u^+$ and $(1)^{1}\Pi_{u}$
states emit UV light at 285.5$-$350.0$\,$nm,
below the $^{1}P$$\rightarrow$${^1S}$ line of the Mg atom at $285.3\,$nm.
The corresponding integrated UV intensity $I_{UV}$ is proportional
to the total population in these molecular states,
reflecting the corresponding PA yield of Mg$_{2}$ molecules.
This population is our control objective.

Experimentally, magnesium vapor of about 5$\,$Torr partial pressure in a
heated cell with argon buffer gas is irradiated by a shaped laser pulse.
The shaping was done using a liquid-crystal spatial light phase
modulator~\cite{WeinerRevSciInst00,BrixnerOptLett01}.
The UV radiation emitted toward the laser-beam entrance to the cell is
collected at a small angle from the laser-beam axis using a proper
optical setup.

Coherent control of femtosecond photoassociation is demonstrated in
Fig.~\ref{fig:scheme}(b) by plotting the UV signal $I_{UV}$, normalized with
respect to the signal obtained for the TL pulse ($k=0$),
versus the chirp parameter $k$. An overall high degree of
chirp control and a strongly asymmetric chirp dependence are observed.
In particular, a large enhancement is
obtained for positively chirped pulses whereas negatively
chirped pulses lead to strong suppression.
The chirp enhancement ($E$) also exhibits an intensity dependence,
which is a clear indication of the strong-field regime:
As the pulse energy, or, equivalently, the peak TL intensity $I_{TL}$, increases,
the maximal enhancement $E_{max}$ and the corresponding chirp $k_{max}$ become
larger. We find $E_{max}$=4.2 at $k_{max}$=0.004$\,$ps$^{2}$ and
$E_{max}$=5.4 at $k_{max}$=0.006$\,$ps$^{2}$
for $I_{TL}=5.0 \times 10^{12}\,$W/cm$^{2}$ and
$7.2 \times 10^{12}\,$W/cm$^{2}$, respectively.
Since two linearly chirped pulses, one of positive chirp $|k|$ and the
other of negative chirp $-|k|$,
have identical instantaneous temporal intensity
but different instantaneous temporal frequency and phase,
the degree of coherent control is best reflected
by the enhancement ratio $E(k_{max}) / E(-k_{max})$.
It amounts here to about 40 for $I_{TL}=7.2 \times 10^{12}\,$W/cm$^{2}$,
i.e., the experimentally observed PA yield is enhanced by this factor
for the positively chirped pulse with $k=k_{max}=0.006\;$ps$^2$
as compared to the negatively chirped pulse.
This striking evidence of phase control calls for an explanation in
terms of the underlying quantum molecular dynamics.

Our first principles modeling of the multiphoton PA process
utilizes the theoretical framework of Ref.~\cite{AmaranJCP13},
combining \textit{ab initio} electronic structure theory  with quantum
molecular dynamics for the Mg$_2$ molecule in the presence of a strong
laser field and thermal averaging based on random phase
wavefunctions.
Here, we extend the model of Ref.~\cite{AmaranJCP13} and
explicitly account for all electronic states shown in
Fig.~\ref{fig:scheme}(a), in order to improve the treatment of the
Stark shifts for the  electronically excited states~\footnote{
  In detail, the Hamiltonian in Eq. (12) of Ref.~\cite{AmaranJCP13}
  has been augmented to include four additional states,
  $(1)^1\Delta_g$ (dissociating into
  $^1D + {^1S}$ atoms), $(1)^1\Delta_u$ ($^1D+ {^1S}$), $(2)^1\Sigma_g^+$ ($^1S+
  {^1S}^\star$), and $(3)^1\Sigma_u^+$ ($^3P+ {^3P}$). If not accounted
  for explicitly, these
  states cause  resonances in the Stark shifts of the
  lower lying states.
}.
The UV emission signal $I_{UV}$
is calculated from the final populations of the
appropriate states, $A^1\Sigma^+_u$, $(2)^1\Sigma^+_u$ and $(1)^1\Pi_u$,
via their Einstein coefficients. These electronic states have
a significant electronic transition dipole moment to the
$X^{1}\Sigma_{g}^{+}$ state and rovibrational levels that are located
below the $^1P + {^1S}$ atomic threshold,
giving rise to emission at wavelengths larger than 285.3$\,$nm.

As Fig.~\ref{fig:scheme}(b) shows, our theoretical model clearly reproduces
the main features of the experimental results -- enhancement of the
signal for positive chirp and suppression for negative chirp. The
dependence on intensity,
i.e., the larger values of $E_{max}$ and $k_{max}$ for larger intensity,
is also predicted qualitatively correctly  by the calculations.
Quantitatively, the simulations show a slightly smaller peak
enhancement, $E_{max}$=4.5 instead of 5.4 for the larger intensity;
and the maximum is located at larger chirps compared to the experimental data.
The discrepancy with respect to $E_{max}$ can easily be
resolved by a small scaling of the $(1)^1\Pi_g$ Stark shift.
For example, scaling this Stark shift by a factor of 0.95,
well within the estimated error bounds of the calculated
polarizabilities, increases $E_{max}$ from 4.5 to 5.8.
On the other hand, the shift in $k_{max}$
is most likely linked to the relative slopes
of the potentials of the $(1)^1\Pi_g$ state and all highly excited states
that are accessed from it.
Due to the number of electronic states that are involved, it is not
possible to identify a single or few parameters whose change would
result in an improved model. The inaccuracy of the highly excited
states of Mg$_2$ in our model is confirmed by recent
spectroscopy~\cite{Tiemann14} which revealed
the well depth of the adiabatic $(1)^1\Pi_u$ state to be larger by
nearly 50\% than the original {\it ab initio}
result~\cite{AmaranJCP13}. This inaccuracy is not surprising:
Potential energy curves of highly excited states are
more prone to error than lower ones since they often
originate from the interaction between two open-shell excited-state atoms.
Such interactions may lead to molecular electronic states which are very different
from the reference ground state and thus demand an even more correlated
approach than the coupled cluster method with single and double excitations
that was employed in Ref.~\cite{AmaranJCP13}.
Moreover, the high density of  electronic states and the occurrence of
possibly numerous avoided crossings between them
result in a multi-reference nature of the electronic problem in the
experimentally probed energy window which cannot be accurately
described by a model based on a single-determinant assumption.
These facts together stretch the capabilities of state of the art \textit{ab initio}
methods. Therefore, more detailed spectroscopic data would be
necessary to improve all
relevant potential energy curves and allow for full quantitative
agreement between theory and experiment.

\begin{figure}[tb]
  \includegraphics[width=\linewidth]{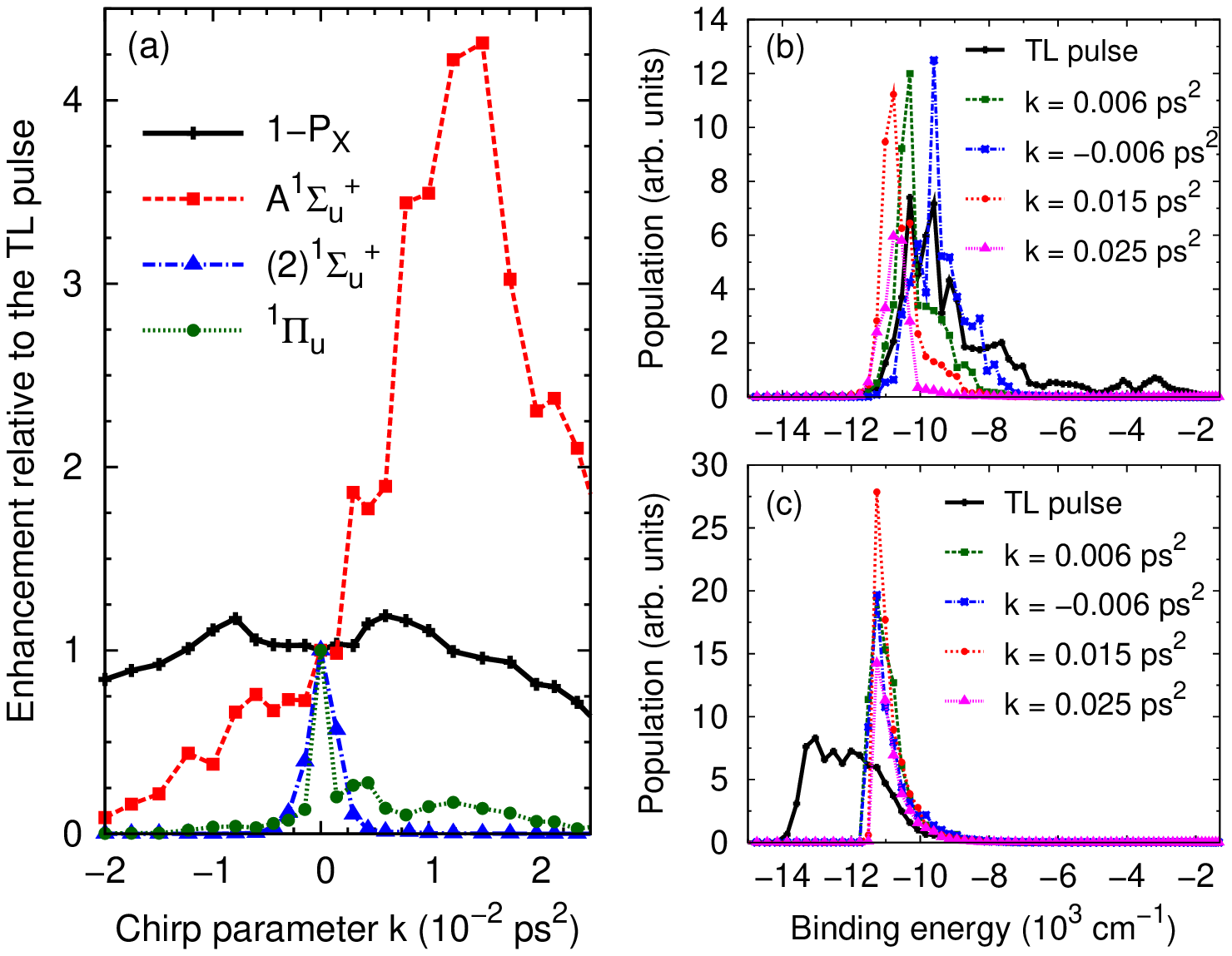}
  \caption{Theoretical results ($I_{TL}=7.2\times 10^{12}\,$W/cm$^2$) for:
    (a) Chirp dependence of the total
    photoassociated population ($1-P_X$)
    and the final populations in the states of the probed UV-emitting band.
    (b) Final vibrational distribution in the intermediate
    $(1)^1\Pi_g$ state for various values of the chirp parameter
    $k$.
    (c) Same as in (b) but obtained within a reduced model
    comprising only the $X^1\Sigma_g^+$ and $(1)^1\Pi_g$ states.
  }
  \label{fig:theo}
\end{figure}
The qualitative agreement observed in Fig.~\ref{fig:scheme}(b) is,
however, certainly sufficient to examine
the theoretical results in view of the
mechanism that underlies the chirp control. To this end,
Fig.~\ref{fig:theo}(a) displays the chirp dependence of all the population
that is photoassociated, given by $1-P_X$, with $P_X$ the final population
in the $X^1\Sigma_g^+$ ground electronic state, comparing it to the
chirp dependence of the final population in the probed UV-emitting states:
Whereas almost no chirp dependence is observed in the total PA yield
($1-P_X$), a clear chirp dependence is seen in the final populations
of the emitting states, in particular a large asymmetry in the population
of the $A^1\Sigma_u^+$ state.
This suggests that the observed chirp dependence does not originate
from the non-resonant $X^{1}\Sigma_{g}^{+}$ to $(1)^{1}\Pi_{g}$ transition,
but rather results mainly from the strong-field dynamics on the
$(1)^1\Pi_g$ and higher lying electronically excited states. If one
assumes the last photon that excites into the UV-emitting states to
constitute a weak probe, it is the shape of the vibrational distribution in
the intermediate $(1)^1\Pi_g$ state that should be responsible for the
chirp dependence of the signal in Fig.~\ref{fig:scheme}(b).
Indeed, the final vibrational distribution in the $(1)^1\Pi_g$ state,
plotted for various chirps in Fig.~\ref{fig:theo}(b), shows a clear
dependence on both sign and magnitude of the chirp parameter.
The analysis is further supported by Fig.~\ref{fig:theo}(c) which
presents, for comparison,
the final vibrational distribution in the $(1)^1\Pi_g$ state,
obtained within a reduced two-state model. It comprises only
the $X^1\Sigma_g^+$ and $(1)^1\Pi_g$ states
rather than all the electronic states of the full model.
The results of the two-state model  differ both
qualitatively and quantitatively from those of the full
model and show, in particular, no dependence on the sign of the chirp.
The chirp dependence in the full model can then be rationalized in
terms of resonant Raman transitions between the $(1)^1\Pi_g$ state
and the higher lying $u$-states: The time dependence of the
instantaneous frequency of the chirped pulse leads to an
up (down) shift of the $(1)^1\Pi_g$ vibrational distribution for
negative (positive) chirp. The magnitude of the up- or down-shift
depends on the absolute value of the chirp parameter.
Down-shifting the $(1)^1\Pi_g$ vibrational
distribution results in an enhanced UV emission signal $I_{UV}$ because it
favors transitions into bound levels of the $A^1\Sigma_u^+$ state in
the probed UV-emitting band, 
whereas an up-shifted $(1)^1\Pi_g$ vibrational distribution is
predominantly excited into dissociative
states which do not contribute to the molecular emission signal.
Our picture of a perturbative final probe photon is confirmed by
comparing the calculated final vibrational distributions in the UV
emitting states to the Franck-Condon projection of the final
$(1)^1\Pi_g$ distribution onto these states.
The narrowing of the vibrational distribution due to the chirp,
observed in both Fig.~\ref{fig:theo}(b) and (c),
is readily understood in terms of a competition between non-resonant
Stark shifts and chirp. The chirp lowers the peak intensity and thus
the Stark shift such that less power broadening of the vibrational
distribution is induced by the chirped pulses as compared to the TL pulse.
Figure~\ref{fig:theo}(b) also indicates why a chirp rate of
$k=0.015\,$ps$^2$ is optimal: For even larger chirps,
we do not observe a further down-shift of the $(1)^1\Pi_g$ vibrational
distribution. This is attributed to both the limited bandwidth of the
pulse and the significantly reduced peak intensity for larger chirps.
Thus the optimal chirp results from a competition between
sufficient intensity for the Raman transitions and
down-shifting of the vibrational distribution.
This interpretation is also supported by the larger
peak enhancement observed for the larger intensity
in Fig.~\ref{fig:scheme}(b).

\begin{figure}[tb]
  \includegraphics[width=\linewidth]{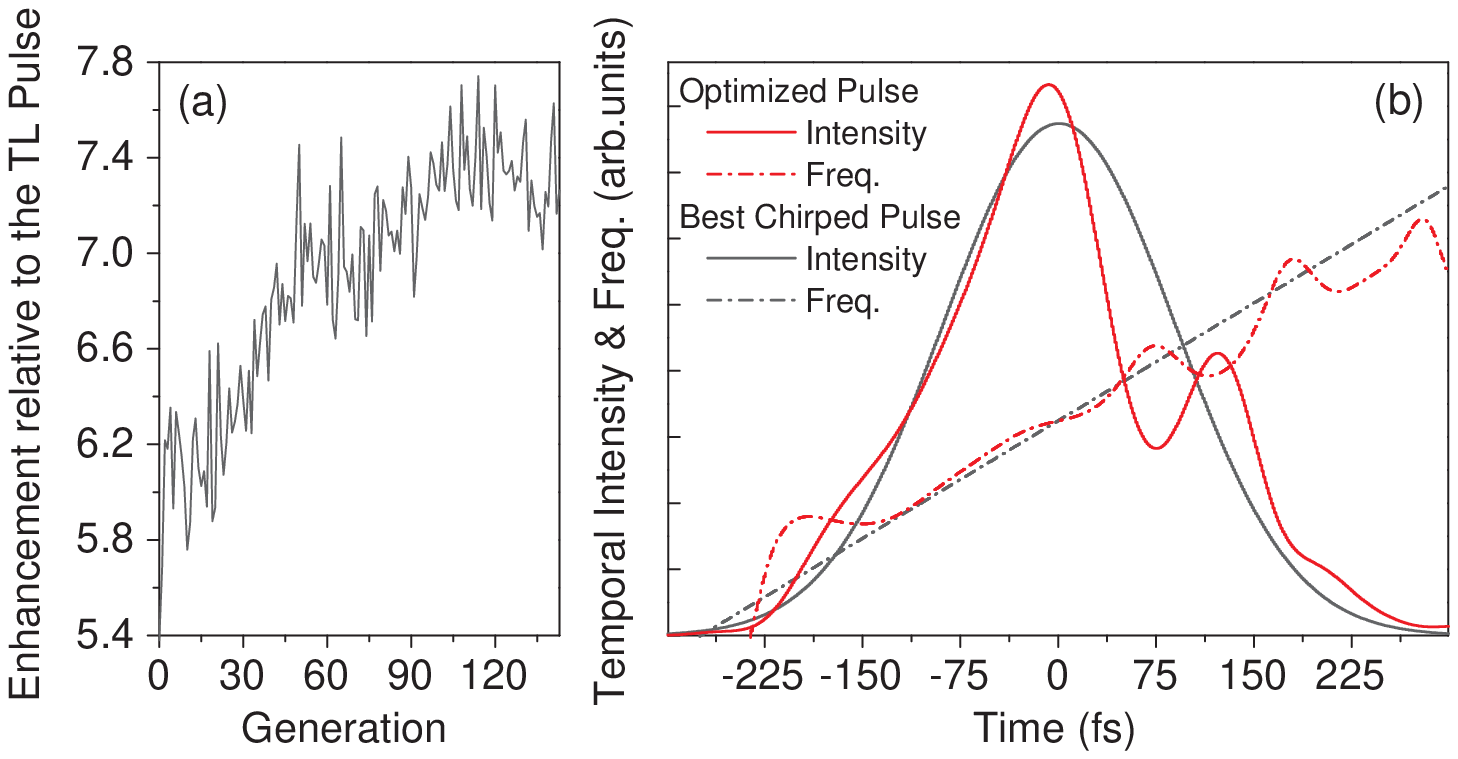}
  \caption{Results for experimental closed-loop phase optimization
    using a learning
    algorithm, starting from the best linearly chirped pulse:
    (a) Enhancement as a function of generation; improvement is very fast.
    (b) The optimized pulse compared to the best linearly chirped
    pulse ($k=0.006\,$ps$^2$).}
  \label{fig:GA}
\end{figure}
Our understanding of the control being facilitated by shaping the
vibrational $(1)^1\Pi_g$ distribution suggests that further
enhancement of the PA signal should be possible by exploiting the
$(1)^1\Pi_g$ vibrational dynamics, in addition to the Raman
transitions.
To explore this possibility, we have experimentally carried out
an automated closed-loop phase optimization using a learning
algorithm~\cite{BrixnerCPC03},
with the measured PA enhancement relative to the TL pulse as the performance
criterion. Each generation of the learning algorithm contains 24
members (chromosomes), where each member is a spectral phase pattern
applied to the pulse. The first generation of the optimization
includes the five best linearly chirped pulses, with
$k=0.004\,$ps to $0.008\,$ps$^{2}$, while all other members are random.
Figure~\ref{fig:GA}(a) shows the maximally obtained PA
enhancement, achieved within each generation, as a function of the
generation number. A fast increase of about 35\% in the maximum
enhancement is observed, from a value of 5.4 at the first generation
to a value of about 7.4 after 130 generations.
The two corresponding pulses, i.e., the best linearly chirped pulse and the
optimized pulse, are shown in Fig.~\ref{fig:GA}(b). While the
optimization essentially keeps the positive linear chirp,
the main change consists in a temporal splitting of the optimized
pulse into two sub-pulses with a time delay of 130$\,$fs.
This time delay corresponds to the vibrational period
of the $(1)^1\Pi_g$ levels in the excitation region.
It indicates that the optimized pulse utilizes
the vibrational dynamics for improving the PA enhancement.

\begin{figure}[tb]
  \centering
  \includegraphics[width=0.95\linewidth]{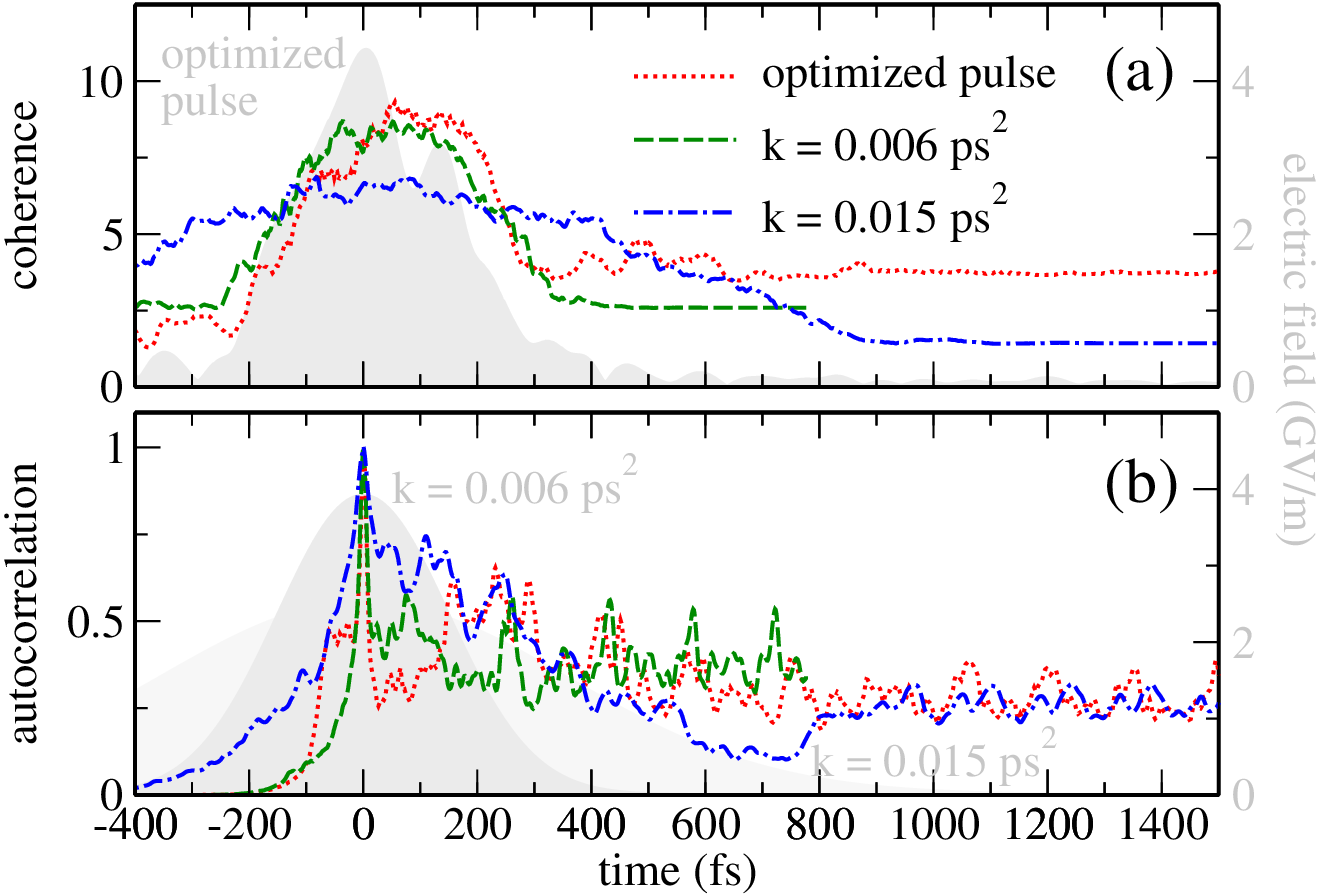}
  \caption{Role of vibrational coherence and dynamics in the $(1)^1\Pi_g$
    state -- vibrational coherence measure (a) and autocorrelation
    function (b).
  }
  \label{fig:dyn}
\end{figure}
When testing the experimentally optimized pulse in our theoretical
model, an enhancement of 7.1  is obtained, surprisingly close to
the experimental value.
Comparing the dynamics under the experimentally optimized pulse
to that of the best linearly chirped pulse reveals the optimized pulse to
populate a significantly broader vibrational band in the $(1)^1\Pi_g$
state, with more population in the lower 
levels that can directly be excited  by a one-photon transition
into the probed UV-emitting band.
The optimized pulse is thus shaped to enhance the transition
probability in the resonant Raman transitions to these 
lower $(1)^1\Pi_g$ levels. The role of coherent vibrational dynamics
in the $(1)^1\Pi_g$ state is further analyzed in Fig.~\ref{fig:dyn} which
displays the vibrational coherence measure~\cite{BaumgratzPRL14},
$\mathcal C(t) = \sum_{i \neq j}|\rho^{\Pi_g}_{ij}(t)|$, and the
autocorrelation function,
$\mathcal A(t)=\mathrm{tr}[\rho^{\Pi_g}(0)\rho^{\Pi_g}(t)]$,
of the normalized $(1)^1\Pi_g$ density. The
values of  $\mathcal C(t)$ in Fig.~\ref{fig:dyn}(a) need to be
compared to the upper bound of the maximally coherent state,
$(d-1)/2$, where $d$ is the number of
levels, about 70 in our case. We thus find a substantial amount of
vibrational coherence in the $(1)^1\Pi_g$ state. In particular at
intermediate times, when the dynamics in the $(1)^1\Pi_g$ state is
relevant, the coherence measure is larger for the optimized pulse than
for the chirped pulses. The second sub-pulse of the optimized pulse
increases the autocorrelation function, reflecting the synchronization
of the pulse delay with the vibrational dynamics.
These observations
confirm that the optimized pulse outperforms the chirped pulses by
utilizing coherent vibrational dynamics in the $(1)^1\Pi_g$ state, in
addition to the chirp-dependent Raman transitions.

In summary, we observe strong-field coherent control of bond
formation in the femtosecond photoassociation of thermally hot magnesium atoms
using phase-shaped laser pulses.
Our modeling from first principles has allowed us to identify
a combination of
Franck-Condon filtering 
in the free-to-bound non-resonant two-photon step with
chirp-dependent resonant Raman transitions and coherent vibrational
dynamics in an intermediate electronic state
to be responsible for the control.
Whereas the purpose of the FC filtering is mainly purification
in order to allow the generation of molecular coherence,
the Raman transitions and vibrational dynamics serve to realize phase
control.
Indeed, the quantum purity in the intermediate state and
final UV-emitting states differ by only 25\% for the experimentally
optimal linear-chirp of $k = 0.006\,$ps$^2$.
Our demonstration of coherent control of bond-making under thermal
conditions points the way toward
controlling transition probabilities and branching ratios to different
target states. For photo-induced chemical reactions with several
product channels, suitable target states would be those that serve  as
a gateway to a different product channel. A feasible route to the
coherent control of photo-induced bimolecular chemical reactions is
now open.

\begin{acknowledgments}
We would like to thank Daniel Reich for technical help at an initial
stage of this work.
This research was supported by The Israel Science Foundation (Grant
No. 1450/10) and the Alexander von Humboldt Foundation (WS).
\end{acknowledgments}


\end{document}